\documentclass[a4paper,12pt]{article}
\pdfoutput=1 

\usepackage{jcappub} 
\usepackage{subfigure}
\usepackage[T1]{fontenc} 

\newcommand{\Fermi}{\textit{Fermi}}

\title{\boldmath Search for right-handed neutrinos from dark matter annihilation with gamma-rays}

\author[a]{Miguel D. Campos,}
\author[a]{Farinaldo S. Queiroz,}
\author[a,b]{Carlos E. Yaguna,}
\author[c]{Christoph Weniger}

\affiliation[a]{Max-Planck-Institut f\"ur Kernphysik, Saupfercheckweg 1, 69117 Heidelberg,
Germany}
\affiliation[b]{Escuela de F\'{i}sica, Universidad Pedag\'ogica y Tecnol\'ogica de Colombia (UPTC)\\
Avenida Central del Norte, Tunja, Colombia}
\affiliation[c]{GRAPPA, Institute of Physics, University of Amsterdam, Science Park 904, 1090 GL Amsterdam, Netherland}
\emailAdd{farinaldo.queiroz@mpi-hd.mpg.de}

\abstract{ 
Several extensions of the Standard Model contain right-handed (sterile) neutrinos in the GeV-TeV mass range. Due to their mixing with the active neutrinos, they may give rise to novel effects  in  cosmology, neutrino physics, and collider searches. In addition, right-handed neutrinos can also appear as final states from dark matter annihilations, with important implications for dark matter indirect detection searches. In this paper, we use current data from the \Fermi\ Large Area Telescope (6-year observation of dwarf spheroidal galaxies) and H.E.S.S.~(10-year observation of the Galactic center) to constrain the annihilation of dark matter into right-handed neutrinos.  We consider right-handed neutrino with masses between 10 GeV and 1 TeV, including both two-body and three-body decays, to 
derive bounds on the dark matter annihilation rate, $\langle \sigma v\rangle$, as a function of the dark matter mass. Our results show, in particular, that the thermal dark matter annihilation cross section, $3\times 10^{-26}\, {\rm cm^3 s^{-1}} $, into right-handed neutrinos is excluded for dark matter masses smaller than $200$ GeV. 
}

\begin{document}
\maketitle
\flushbottom

\section{Introduction}
\label{sec:intro}

The existence of dark matter in our universe has been established for different times and distance scales of the universe. Its fundamental nature is one of the most important open problems in science \cite{Strigari:2013iaa,Profumo:2013yn,Bertone:2016nfn,Queiroz:2016sxf,Queiroz:2016awc}. From a particle physics perspective, dark matter is often interpreted as Weakly Interacting Massive Particles (WIMPs), which feature electroweak scale interactions predicting signals within current and planned experiments. 
Among the search strategies for WIMPs, indirect dark matter detection has the advantage of connecting particle physics with astrophysics.  The gamma-ray signal from dark matter annihilation in our galaxy, for instance, depends on  the dark matter density profile, the mass of the dark matter particle, and its self-annihilation cross section into different final states. The former is purely an astrophysical quantity, inferred through either cosmological simulations or galaxy rotation curves, whereas the others are particle physics inputs that vary according to the specific model of dark matter considered. \\

As far as indirect dark matter detection is concerned, the final states from dark matter annihilation are often assumed to belong to the Standard Model spectrum, such as charged leptons, quarks, gauge bosons and neutrinos \cite{Calore:2013yia,Bringmann:2011ye,Hooper:2012sr,Ibarra:2013zia,Bergstrom:2013jra,Lin:2014vja,Bringmann:2014lpa,Gonzalez-Morales:2014eaa,DiMauro:2015jxa,Mambrini:2015sia,Buckley:2015doa,Giesen:2015ufa,Lu:2015pta,Cavasonza:2016qem,Belotsky:2016tja,Profumo:2016idl,Huang:2016tfo,Hooper:2016cld,Caputo:2016ryl,Jin:2017iwg}\footnote{Some exceptions occurs for annihilations into metastable particles which then decay into SM particles \cite{Mardon:2009rc,Abdullah:2014lla,Cerdeno:2015ega,Martin:2014sxa,Dutta:2015ysa,Rajaraman:2015xka}}. In this work,  we investigate instead the annihilation of dark matter into right-handed neutrinos.  \\

Right-handed neutrinos naturally appear in several well-motivated extensions of the Standard Model, and they are essential for the generation of neutrino masses through the type-I see-saw mechanism  \cite{Minkowski:1977sc,Mohapatra:1979ia,Lazarides:1980nt,Mohapatra:1980yp,Schechter:1980gr}. In addition, they are usually invoked to account for the matter-antimatter asymmetry via leptogenesis \cite{Abada:2015rta,DiBari:2016guw},  they may induce interesting collider signatures \cite{Tello:2010am,Melfo:2011nx,Nemevsek:2011hz,Anchordoqui:2012qu,CMS:2012zv,Lindner:2013awa,Khachatryan:2014dka,Deppisch:2015qwa,Ng:2015hba,Antusch:2016ejd,Das:2017nvm}, and sometimes they are even considered as dark matter candidates \cite{Kaneta:2016vkq,Okada:2016tci,Dev:2016xcp,Heurtier:2016iac}. Hence, the presence of right-handed neutrinos is common to several scenarios with a rather rich phenomenology. Moreover, it has been recently argued that, within the minimal seesaw scenario, the masses of the right-handed neutrinos should lie below $10^3$ TeV \cite{Bambhaniya:2016rbb}, a mass range that widely overlaps with that expected for WIMPs.  It is therefore possible that the dark matter particle annihilates into right-handed neutrinos. In fact,  right-handed neutrinos appear as  final states from dark matter annihilation in several explicit models considered previously in the literature (see for example \cite{Allahverdi:2009se,Queiroz:2014yna,Dudas:2014ixa,Tang:2015coo,Allahverdi:2016fvl,Ibarra:2016fco,Alves:2016fqe,Garcia-Cely:2017oco} and references therein), and, in some cases \cite{Mizukoshi:2010ky,Alvares:2012qv,Profumo:2013sca,Kelso:2014qka,Dong:2014wsa,Cogollo:2014jia,Tang:2015coo,Rodejohann:2015lca,Alves:2015mua,Klasen:2016qux,Escudero:2016tzx,Escudero:2016ksa}, they  constitute the  dominant annihilation mode of the dark matter.\\

Here, we use state-of-art techniques and up-to-date data from \Fermi-LAT and H.E.S.S.~to constrain the annihilation of dark matter into right-handed neutrinos. The rest of the paper is organized as follows: the next section describes the basic framework of our analysis; in section \ref{sec:dmann} the energy spectrum expected from dark matter annihilation into right-handed neutrinos is analyzed; sections \ref{sec:fermi} and \ref{sec:hess} specify the data sets and the methods used to derive the constraints, which are our main results and are presented in section \ref{sec:results}; finally, our results are summarized in section \ref{sec:con}.

\section{Framework}
 
The search for right-handed neutrinos from dark matter annihilation relies on their interactions with SM particles. In this work, we will assume they interact with the SM particles via the see-saw mechanism type-I, i.e.~their
interactions with SM particles occurs via the mixing with the active neutrinos. Thus, right-handed neutrinos produced from dark matter annihilation have decay modes with the presence of either gauge bosons ($W^{\pm}\, l, Z \nu$) or the Higgs ($H\,\nu$), which then decay producing quarks and leptons. The quarks and leptons resulted from the decay cascade yield,  through hadronization or final state radiation processes, gamma-rays that can then be detected using \Fermi-LAT and H.E.S.S.~telescopes. \\

\begin{figure}[!h]
\centering
\includegraphics[width=0.3\columnwidth]{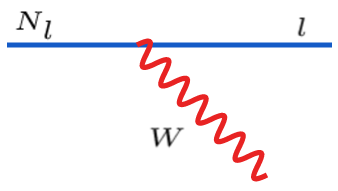}
\includegraphics[width=0.3\columnwidth]{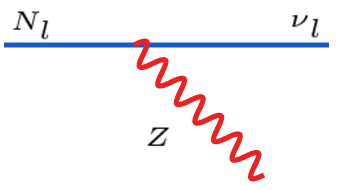}
\includegraphics[width=0.3\columnwidth]{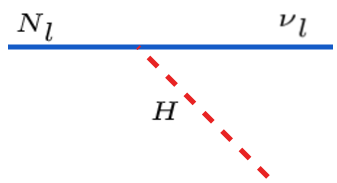}
\caption{Feynman diagrams for two-body decays of the right-handed neutrinos. Notice that we consider the one flavor approximation, where the lepton produced having the same flavor of the right-handed neutrino. The three body decays follow straightforwardly from W,Z and H decays. See text for details.}
\label{Ndecays}
\end{figure}

In order to precisely compute the amount of gamma-rays generated in these processes we first need to account for the right-handed neutrino decays. The rates of two-body decays into $W^{\pm}l$, $Z\nu$ and $H\nu$ are as follow (see Fig.\ref{Ndecays}),

\begin{equation}
\begin{aligned}
\Gamma (N_l\to W^\pm\ell^\mp) & = \dfrac{\alpha_W}{16 M_W^2}|B_{lN}|^2M_N^3\left(1+\dfrac{2M_W^2}{M_N^2}\right)
\left(1-\dfrac{M_W^2}{M_N^2}\right)^2 \theta(M_N-M_W), \\
\Gamma (N_l\to Z\nu_\ell) & = \dfrac{\alpha_W}{16 M_W^2}|C_{\nu N}|^2 M_N^3\left(1+\dfrac{2M_Z^2}{M_N^2}\right)
\left(1-\dfrac{M_Z^2}{M_N^2}\right)^2  \theta(M_N-M_Z), \\
\Gamma (N_l\to H\nu_\ell) & = \dfrac{\alpha_W}{16 M_W^2}|C_{\nu N}|^2 M_N^3\left(1-\dfrac{M_H^2}{M_N^2}\right)^2  \theta(M_N-M_H),
\end{aligned}
\label{Eq:1}
\end{equation} where $\alpha_W$ is $g^2/4\pi$ and $B_{lN} (C_{\nu N})$ represent the  mixing matrices entering the charged (neutral) current in agreement with \cite{Buchmuller:1990vh,Pilaftsis:1991ug}. We assume that dark matter is its own anti-particle, which is causing the factor $1/2$.
As for the three body final state decays, relevant  whenever $M_N < M_W,M_Z$, we used the expressions appearing in \cite{Gorbunov:2007ak,Atre:2009rg,Shuve:2016muy}.

For completeness, we also included the radiative decay  into an active neutrino and a photon -- see e.g.~\cite{Kusenko:2009up}. All two- and three-body decays were implemented in \textsc{Pythia 8.219}, taking into account both leptonic and hadronic decay modes.\\

In our analysis, we consider the so-called `one-flavor approximation', which means that the right-handed neutrino mixes only with one lepton flavor at a time. As a result, $B_{lN} =C_{\nu N}$ and the decay branching ratios of the right-handed neutrino, which are the relevant quantities for indirect detection purposes, depend only on its mass --not on  the mixing structure. Let us emphasize, in any case,  that this approximation is not expected to significantly affect our results because the lepton flavor in the final state is not that relevant, as we shall see further. Thus, the inclusion of all flavors would simply bring unnecessary complications yielding rather mild quantitative changes in our results.

\medskip

Thus, in contrast to collider or laboratory searches, which strongly depend on the mixing  of the right-handed neutrinos, the indirect detection signal from dark matter annihilation is practically insensitive to it. Indirect dark matter detection introduces, therefore, a new way to test some particle physics models in the context of the type-I seesaw mechanism even in the regime of suppressed mixings.\\

Now that we have reviewed the framework on which our results are based, we discuss in detail the signal from dark matter annihilation we are interested in.

\begin{figure}[!h]
\centering
\includegraphics[width=0.3\columnwidth]{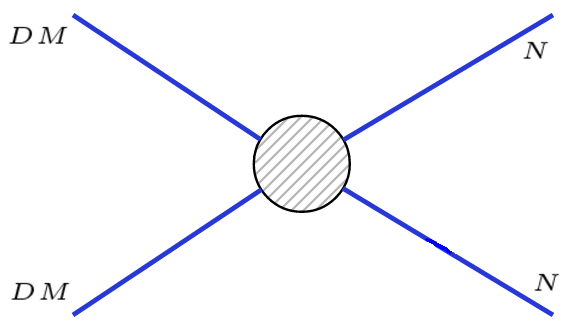}
\caption{Feynman diagrams for dark matter annihilation into right-handed neutrinos.}
\label{fig:annihilation}
\end{figure}

\begin{figure}[!h]
\centering
\includegraphics[width=0.45\columnwidth]{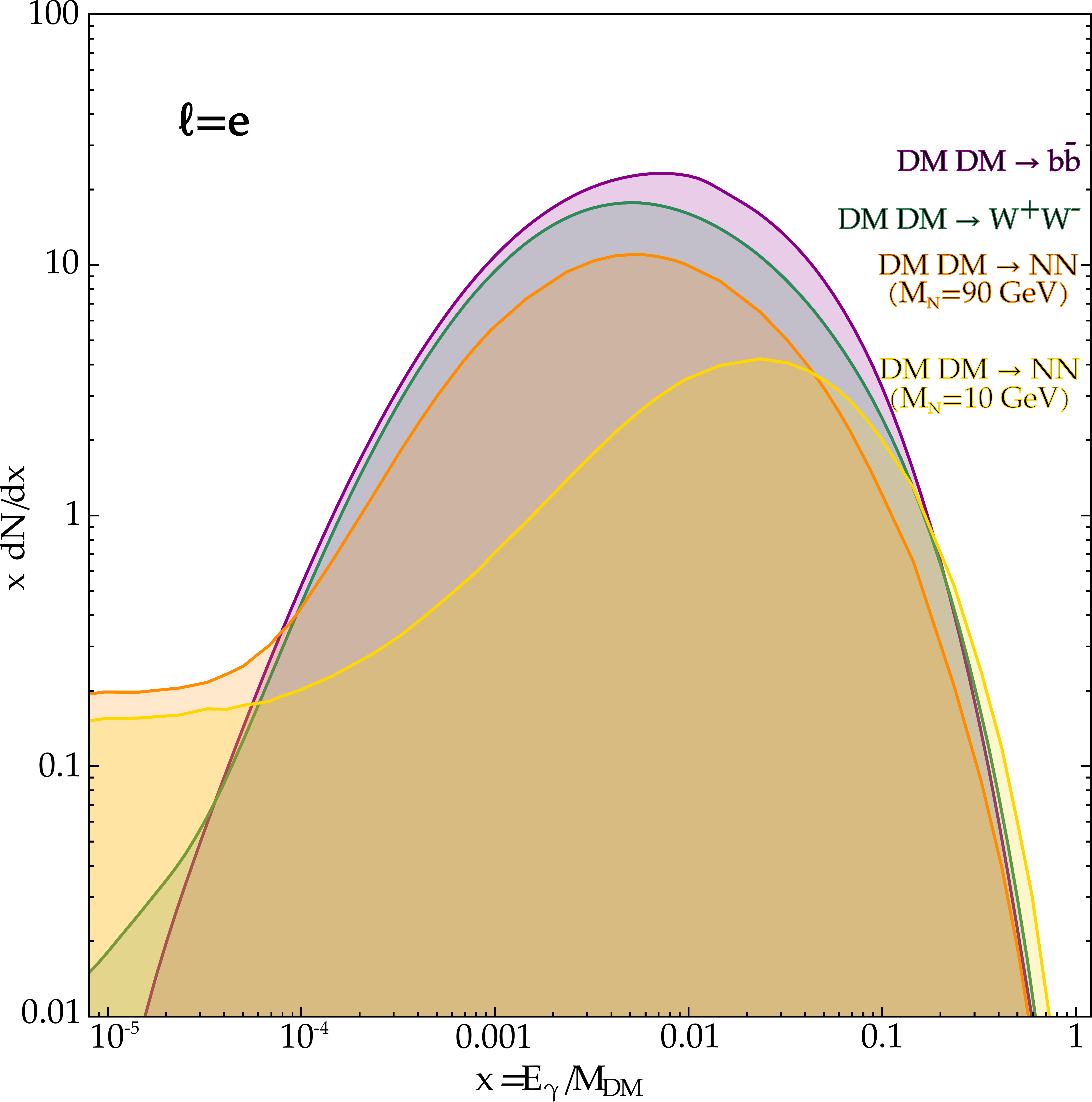}
\includegraphics[width=0.45\columnwidth]{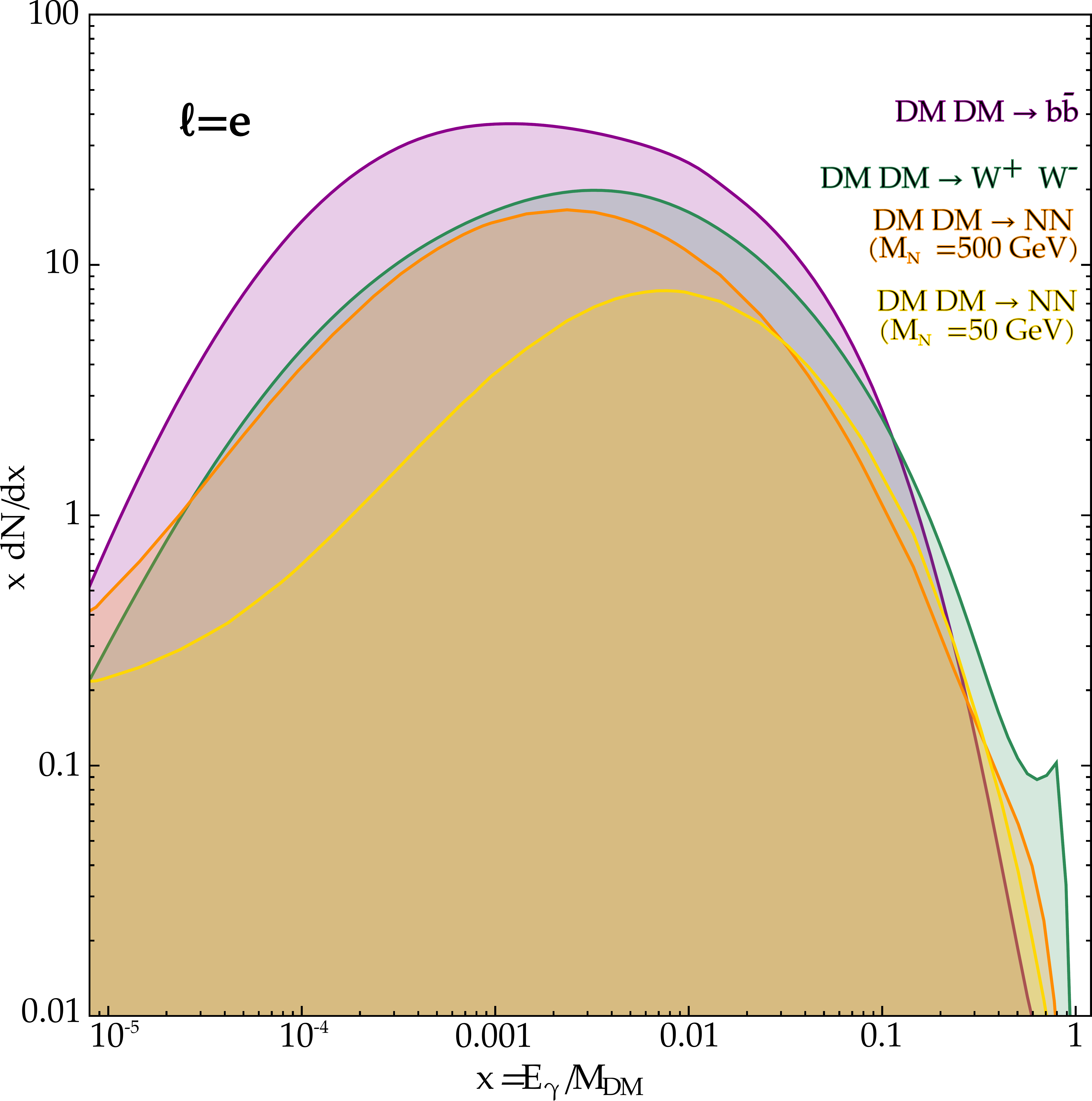}
\caption{\label{fig:xdNdx1} 
Gamma spectra for the annihilation of a DM particle with mass $M_{\mathrm{DM}}=100$ ~GeV (left-panel) and $M_{\mathrm{DM}}=1000$~GeV (right panel) into different final states. For simplicity we show just the case in which $N$ mixes exclusively with $\ell=e$. Magenta and green curves represent the final states of $b\bar{b}$ and $W^+W^-$. {\it Left-Panel:} Orange and yellow contours account for annihilations into right-handed neutrinos with $M_N=90$ GeV and $M_N=10$ GeV respectively.  {\it Right-Panel:} Orange and yellow contours account for annihilations into right-handed neutrinos with $M_N=500$ GeV and $M_N=50$ GeV respectively.}
\end{figure}

\begin{figure}[!h]
\centering
\includegraphics[width=0.6\columnwidth]{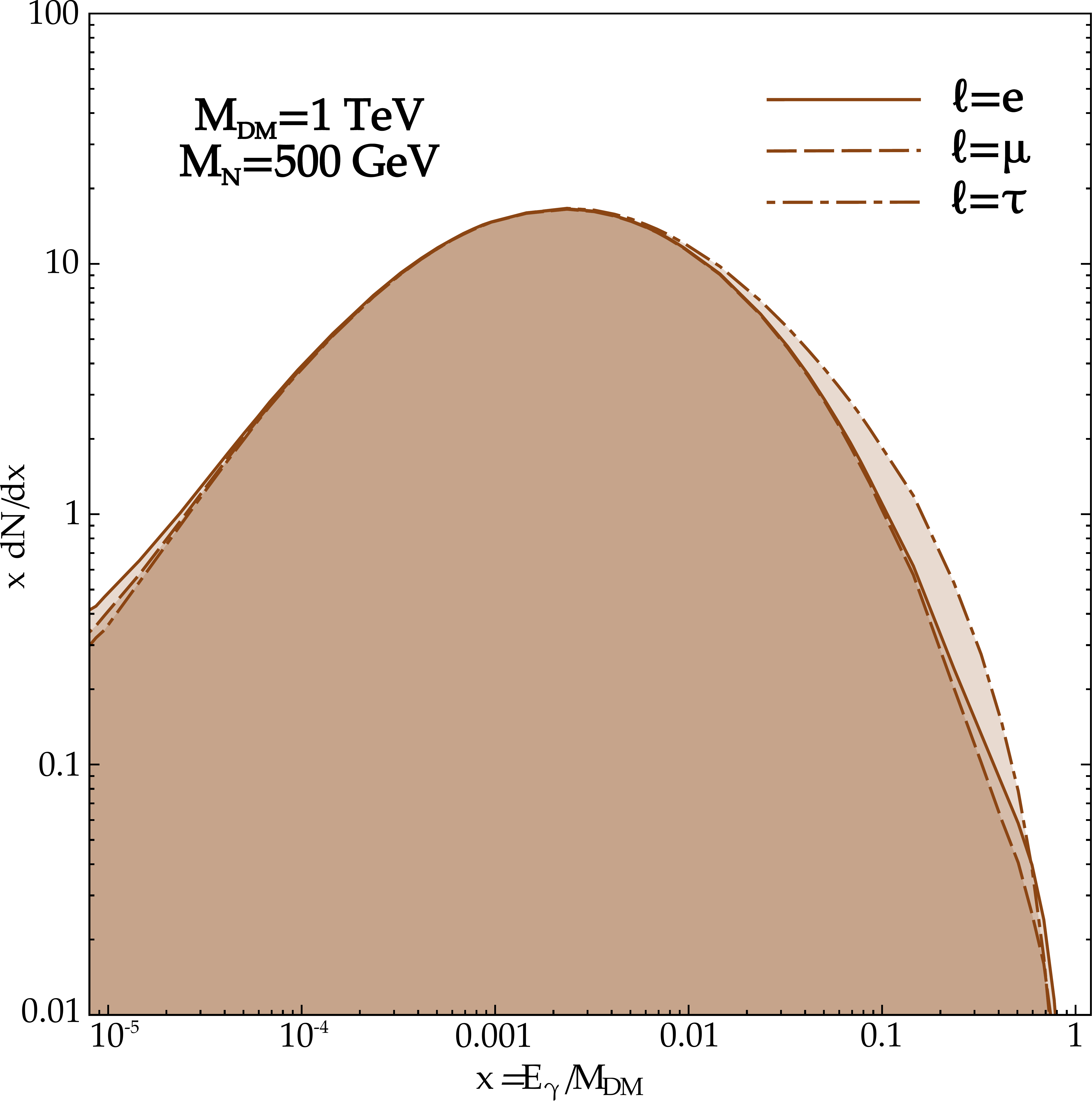}
\caption{\label{fig:xdNdx2} 
The energy spectrum for $M_{DM}=1$~TeV and $M_N=500$~GeV, for different final state leptons. One can see there is only a mild different between them, with the final state $\tau$ leading to harder gamma-ray yield, i.e. larger $dN/dE$, as expected since they lead to a relatively more efficient hadronization process.}
\end{figure}

\section{Dark Matter Annihilation}
\label{sec:dmann}

Since the presence of dark matter has been established, it is expected  that these dark matter particles may self-annihilate in regions where the concentration of dark matter is large enough, as is the case in the Galactic Center and in dSphs. A gamma-ray indirect dark matter detection signal is translated at the end of the day to counting the number of photons at a given energy bin from a portion of the sky, and the quantity that captures this information is the differential flux. Since we are dealing with annihilation, the differential flux should be proportional to: the number density squared $n_{\chi}^2=\rho_{DM}^2/M_{DM}^2$, where $\rho_{DM}$ and $M_{DM}$ are the dark matter density and mass of the dark matter particle; the annihilation cross section times velocity ($\sigma v$); the number of photons produced per dark matter annihilation as a function of energy, i.e the energy spectrum ($dN/dE$); and finally the size and density of the region of the sky under study, enclosed in the so-called J-factor ($J_{ann}$). After including the normalization factor of $4\pi$, which accounts for the solid angle of a sphere, we find the differential flux for dark matter annihilation within an angular direction to be,

\begin{equation}
\label{eq:flux}
\frac{d\Phi_\gamma(\Delta \Omega)}{dE}(E_\gamma)=\frac{1}{4\pi}\frac{\sigma v}{2 M_{DM}^2}\frac{dN_\gamma}{dE_{\gamma}}\cdot J_{ann}
\end{equation}
where $J_{ann}$ is the annihilation J-factor, 
\begin{equation}
   J_{ann} \; =\; \int_{\Delta \Omega} d\Omega \int \rho_{DM}^2 (s) ds\;,
 \label{eq:Jfac_def}
\end{equation} with $s\; =\; s(\theta )$, and the integral is computed over the line of sight within the solid angle and the factor of two in Eq.\eqref{eq:flux} appears because we assuming the dark matter particle to be its own anti-particle.\\

The dark matter density profile that goes in the computation of the J-factors is not precisely determined \cite{Burkert:1995yz,Salucci:2000ps,Graham:2005xx,Salucci:2007tm}. The \Fermi-LAT collaboration adopted the ones based on the Navarro-Frenk and White  (NFW)  profile, which is motivated by the results of numerical simulations that include gravitational forces only \cite{Navarro:2008kc}  \footnote{Keep in mind that steeper profiles seem to be favored by numerical simulations that include the effects of the existence of baryons \cite{Prada:2004pi,Gnedin:2004cx,Seigar:2014kta,Mollitor:2014ara,Zhu:2015jwa,Calore:2015oya,Chan:2015tna,Hooper:2016ggc,Gammaldi:2016uhg,Peirani:2016qvp,Peters:2016qjm,Suto:2016zqb,Peters:2016yeq,Richards:2016lxz,Contreras:2016hdu,Reischke:2016dop,Hayashi:2016kcy,Vega:2016bcn,Reischke:2016eza,Pace:2016oim,Merten:2017jpf,Rodrigues:2017vto}.}. For this reason we will stick to it, using

\begin{equation}
   \rho_{DM} (r) \; =\; \frac{\rho_s}{r/r_s (1+ r/r_s)^2}\;,
 \label{eq:DM_profile}
\end{equation}where $r_s$ and $\rho_s$ are the scale radius and the characteristic density respectively.\\

For dSphs these inputs are determined dynamically from the maximum circular velocity $v_c$ and the enclosed mass contained up to the radius of maximum $v_c$  as discussed in \cite{Ackermann:2013yva}.  In this work we adopted the J-factors listed in the table I of \cite{Ackermann:2015zua} which is equivalent to \cite{Ackermann:2013yva}. As for the galactic center, the dark matter density profile is extracted from simulations of Milky Way-like halos, and the density profile favored  by these simulations also varies. In our work we adopt the values quoted by the H.E.S.S.~collaboration in  \cite{Abramowski:2011hc}, which can also be computed with the code CLUMPY \cite{Bonnivard:2015pia}.\\ 

Lastly, the energy spectrum in Eq.\eqref{eq:flux} was obtained numerically with \textsc{Pythia 8.219} by simulating a resonance $\mathcal{D}$ with $M_\mathcal{D}=2M_{\mathrm{DM}}$ that decays into two right-handed neutrinos,  as done in \cite{Cirelli:2010xx} (see Fig.\ref{fig:annihilation}). We investigated right-handed neutrino masses in the range $[10\text{GeV},1\text{TeV}]$ for which we needed to consider explicitly 3-body decays when $M_N < M_W,M_Z$. Both two-body and three-body decays obtained in  \cite{Gorbunov:2007ak,Atre:2009rg,Shuve:2016muy} were coded into 
\textsc{Pythia 8.219} to numerically to obtain the energy spectrum as shown in Fig.\ref{fig:xdNdx1}. We did not include inverse Compton scattering. This indeed could improve even further our limits specially for the electron and $\mu$ right-handed neutrino flavors since would increase the number of photons at higher energies.  Thus, our analysis is conservative from this perspective.

\medskip

Notice that, regardless of the decay mode relevant for a given right-handed neutrino mass, both leptonic and hadronic decays produce a sizable amount of gamma-rays, which can be detected by \Fermi-LAT and H.E.S.S.~instruments. In other words, the signature of dark matter annihilation into right-handed neutrinos is the usual continuous gamma-ray emission. To clearly see that, in  Fig.\ref{fig:xdNdx1} we exhibit the energy spectrum for a dark matter mass fixed at $100$~GeV ({\it left-panel}) and $1000$~GeV ({\it right-panel}), but for different right-handed neutrino masses. \\

We superimpose our results with the energy spectrum of benchmark final states used in the literature, namely $b\bar{b}$ and $W^+W^-$, obtained with the PPPC4DMID code \cite{Cirelli:2010xx} (including  electroweak corrections). In the {\it left-panel} of Fig.~\ref{fig:xdNdx1}, the magenta, green, orange and yellow curves represent final states of $b\bar{b}$, $W^+W^-$, $NN$ with a mass of $M_N=90$ GeV, and $NN$ with a mass of $M_N=10$ GeV respectively. In the {\it right-panel}, the orange and yellow contours now account for the energy spectrum with right-handed neutrino masses of $500$~GeV and $50$~GeV respectively.\\

Note that when the right-handed neutrino is heavy enough to decay into on-shell gauge W and Z gauge bosons, see {\it left-panel}, the energy spectrum for annihilation into right-handed neutrinos, though softer, resembles the $b\bar{b}$ and $W^{+}W^{-}$ ones at higher energies, but at lower energies it yields a harder gamma-ray spectrum, i.e. larger $dN/dE$, due to the presence of leptonic decay modes which produce much more photons at lower energies compared to hadronic ones. From this fact only, one can already foresee that the \Fermi-LAT and H.E.S.S.~limits on the dark matter annihilation into right-handed neutrinos should be comparable to the bounds based on these canonical channels. Moreover, as we ramp up the dark matter to $1$~TeV as shown in the {\it right-panel}, the energy spectrum becomes visibly harder at lower energies for both $M_N=500$~GeV and $50$~GeV masses. With these results at hand, we next discuss the data set used in this work.  

\label{s:res}
\begin{figure}[!t]
\centering
\includegraphics[width=0.9\columnwidth]{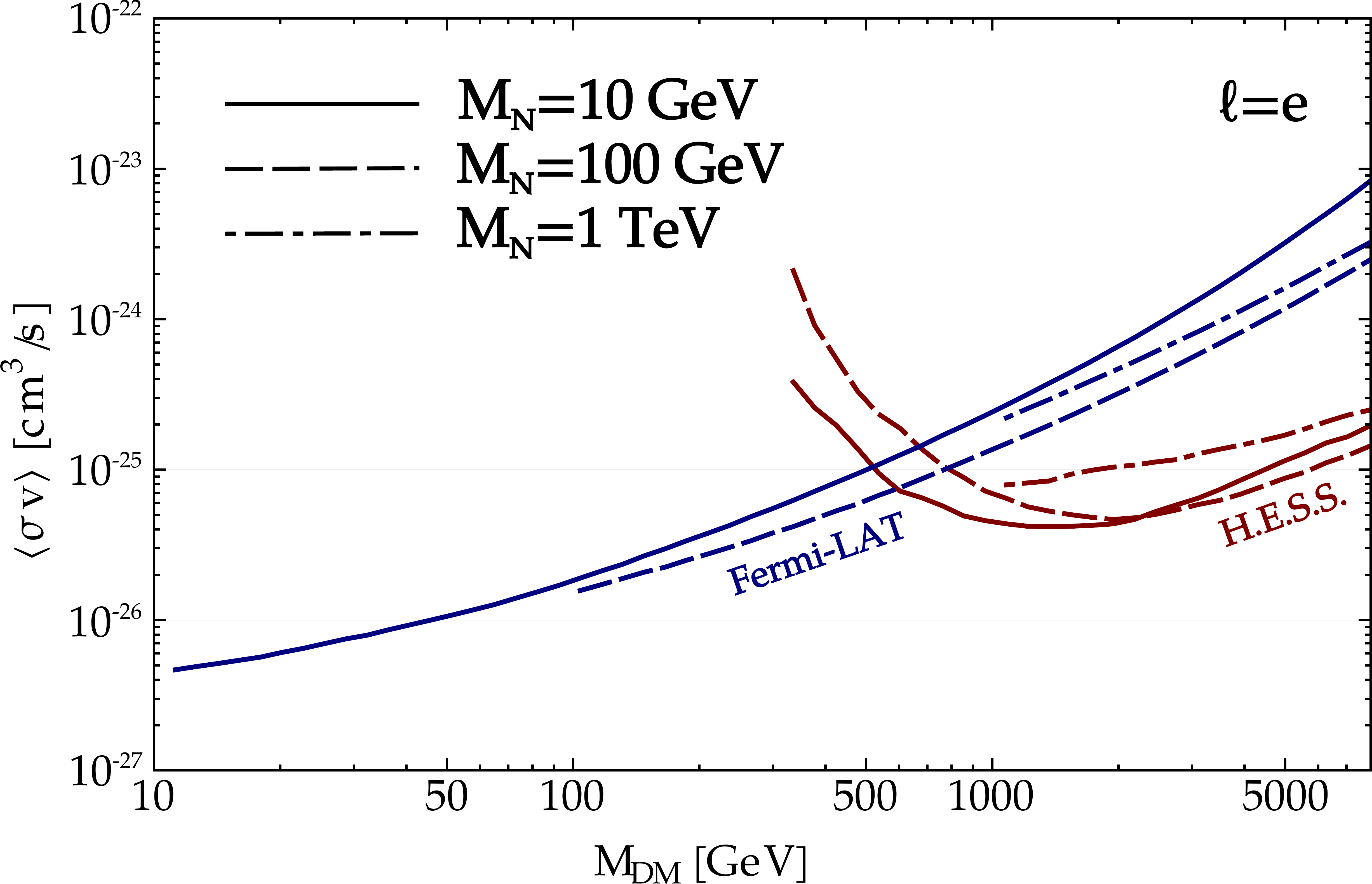}
\caption{\label{limit1} 
Upper limit on $\langle \sigma v \rangle$ as a function of the dark matter mass ($M_{DM}$) for the electron right-handed neutrino final state. Blue curves represent the limits obtained using \Fermi-LAT data, while red are those obtained using H.E.S.S.~data.
Solid, dashed and dot-dashed curves are for $M_N=10,10^2,10^3$ GeV respectively.}
\end{figure}

\section{Fermi-LAT limits}
\label{sec:fermi}

The NASA telescope \Fermi-LAT has given rise to a new era of gamma-ray analysis due the publicly available data and the effort in developing friendly user tools  which allowed a multitude of independent assessments and cross-checks of results, specially in the dark matter field  \cite{Ahnen:2016qkx,Oman:2016zjn,Li:2015kag,Zitzer:2015eqa,
Rico:2015nya,Bonnivard:2015tta,Bonnivard:2015xpq,Hooper:2015ula,Geringer-Sameth:2015lua,Drlica-Wagner:2015xua,Baring:2015sza,Queiroz:2016zwd}, in addition to major improvements in the detector performance compared to its predecessors.\\

DSphs are dark matter dominated objects with a relatively known dark matter density profile, and thanks to their high latitude they suffer from low diffuse gamma-ray emission. For these reasons dSphs are recognized as standard targets in the search for a dark matter signal. Albeit, there are still uncertainties surrounding the dark matter content in dSphs, and for this reason a combined data set of dSphs allied with a maximum likelihood analysis was adopted in order to make more robust assessments concerning a potential dark matter signal and limits. That said, the \Fermi-LAT telescope has been collecting data from dozens of dSphs in the Milky Way. They have gathered six years of data, in particular photons with energies between $500$~MeV and $500$~GeV belonging to the event class P8R2SOURCEV6, and using up-to-date software, PASS-8. The new analysis improves upon previous studies in several aspects, namely improved effective area, point-spread function, updated background models, and the inclusion of the latest gamma-ray catalog.  Since no significant gamma-ray excess was observed in these years of observations, restrictive constraints were placed on the dark matter properties \cite{Ackermann:2015zua}.

\medskip

As we have seen in the previous section, once the astrophysical quantities
are defined, a dark matter signal depends on three particle physics quantities, namely the annihilation cross section, energy spectrum and dark matter mass.
Thus after selecting the annihilation final state, the collaboration was able to show the resulting bounds on the plane {\it annihilation cross section vs dark matter mass} for $b\bar{b}$, $W^+W^-$ among other SM particles \cite{Ackermann:2015zua} by performing a joint likelihood analysis of a stack of dSphs where the statistical errors in the J-factors were marginalized over. \\

\begin{figure}[h]
\centering
\subfigure{
\includegraphics[width=0.9\columnwidth]{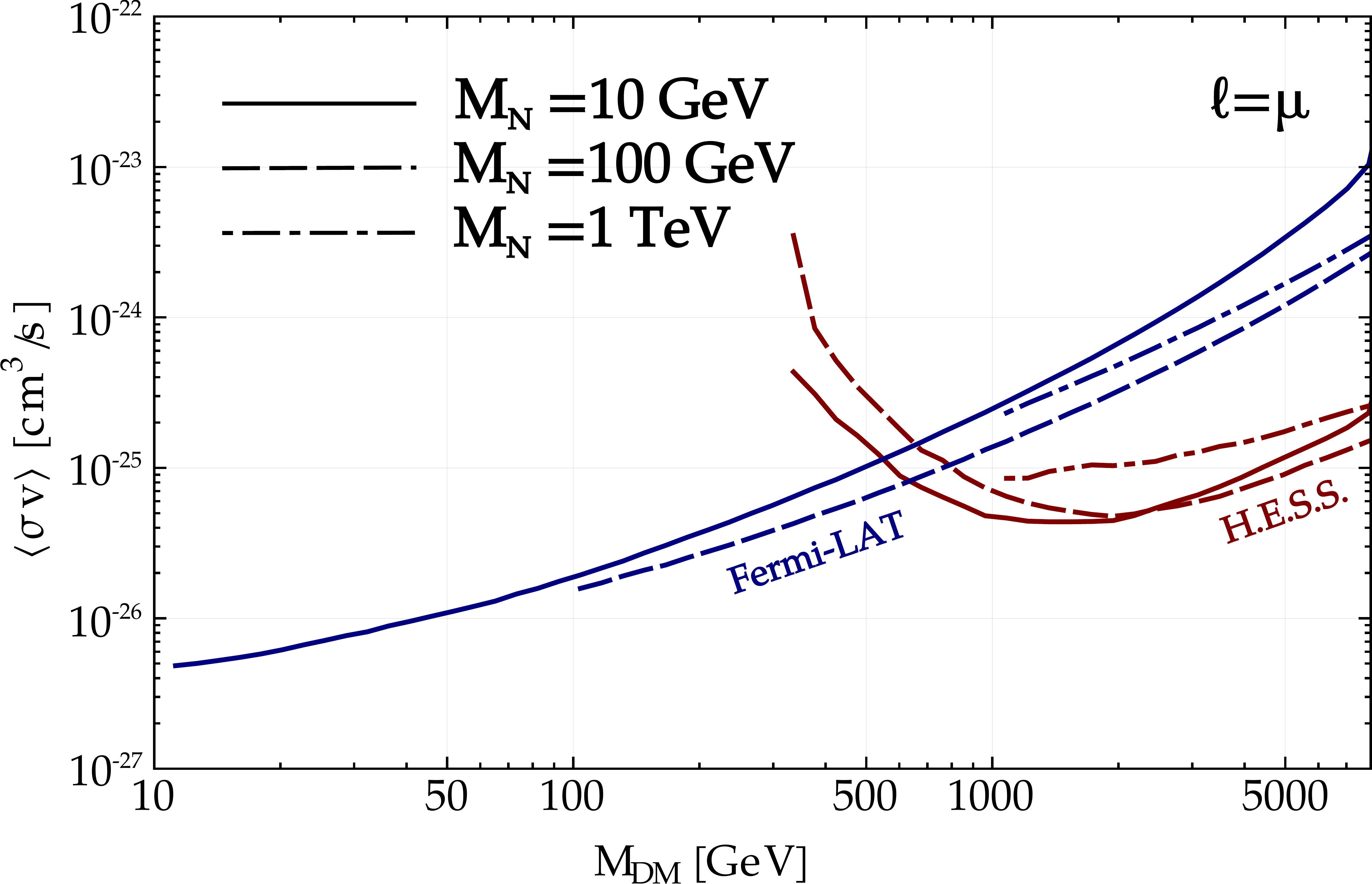}}
\caption{\label{limit2} 
Upper limit on $\langle \sigma v \rangle$ as a function of the dark matter mass ($M_{DM}$) for the muon right-handed neutrino final state.
Blue curves represent the limits obtained using \Fermi-LAT data, while red are those obtained using H.E.S.S.~data.
Solid, dashed and dot-dashed curves are for $M_N=10,10^2,10^3$ GeV respectively.
}
\end{figure}

\begin{figure}[h]
\centering
\subfigure{
\includegraphics[width=0.9\columnwidth]{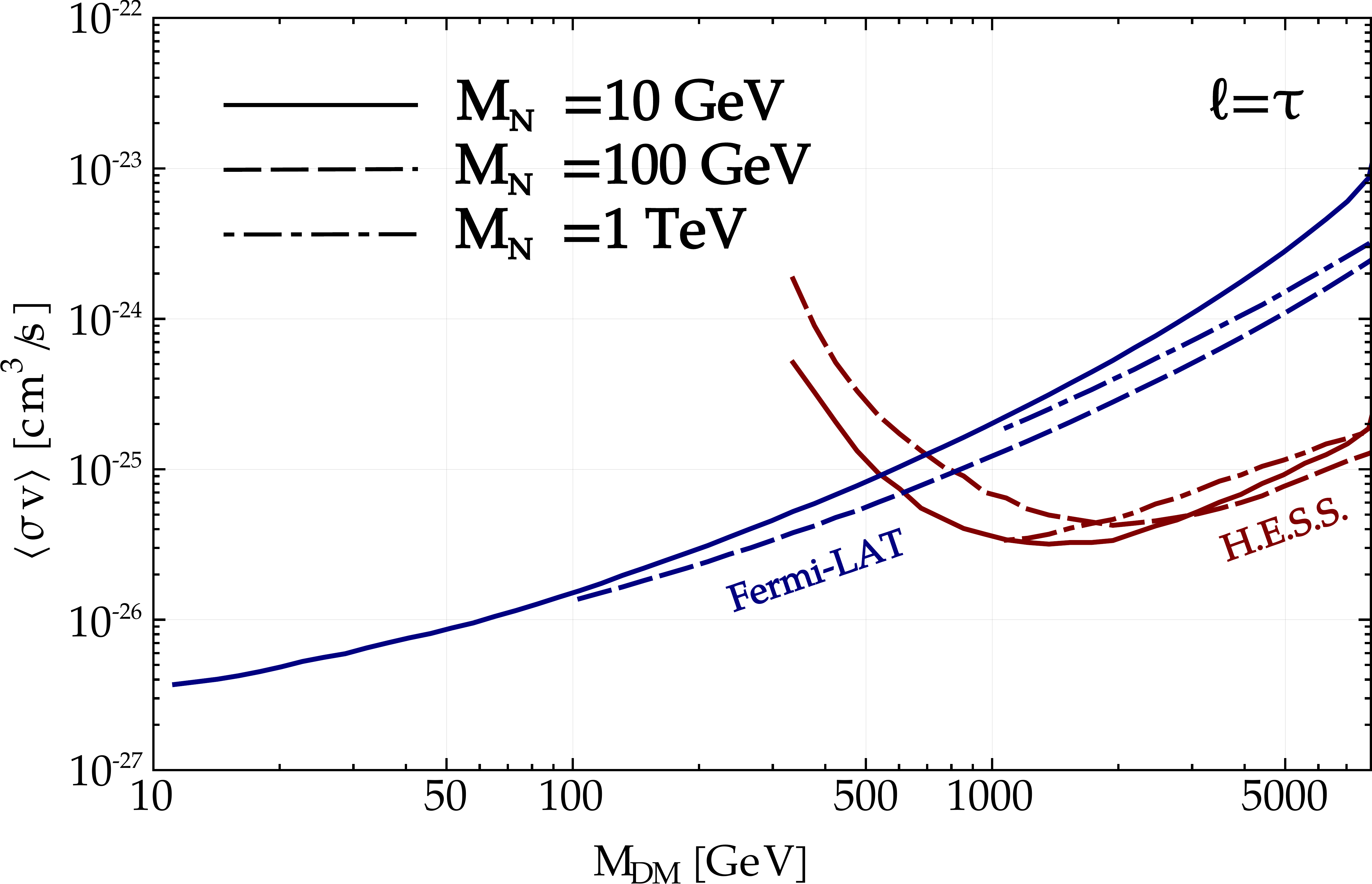}}
\caption{\label{limit3} 
Upper limit on $\langle \sigma v \rangle$ as a function of the dark matter mass ($M_{DM}$) for tau right-handed neutrino final state. Blue curves represent the limits obtained using \Fermi-LAT data, while red are those obtained using H.E.S.S.~data.
Solid, dashed and dot-dashed curves are for $M_N=10,10^2,10^3$ GeV respectively.
}
\end{figure}

The \Fermi-LAT team has made publicly available,  however, only the energy binned Poisson likelihood tools of each individual dSph \footnote{http://www-glast.stanford.edu/pub\_data/1048/}, which allows one to reproduce the limit originated from a given dSph. In order to recast their limits in a more solid way, and not be sensitive to peculiarities of a particular dSph, a joint-likelihood study is necessary. To do so, we follow the supplemental material of Ref.~\cite{Ackermann:2015zua}, and perform a joint-likelihood analysis across 15 dSphs, and treat the J-factors as nuisance parameters, by defining a J-factor likelihood function as follows,
\begin{eqnarray}
\mathcal{L}_J (J_i| J_{obs,i}, \sigma_i) &= & \frac{1}{\ln(10)J_{obs,i}\sqrt{2\pi} \sigma_i}\nonumber\\
& \times  & \exp \left\{ -\frac{(\log_{10}(J_i)-\log_{10}(J_{obs,i}))^2}{2\sigma_i^2} \right\}\nonumber\;,
\end{eqnarray}
where $J_{obs,i}$ is the measured J-factor with error $\sigma_i$ of a dSph $i$ and $J_i$ is the true J-factor value. In this way, the joint-likelihood analysis, i.e. the overall product of each individual likelihood, is performed by computing $\mathcal{L}_i (\mu, \theta_{i}| D_i)$ as below,,
\begin{eqnarray}
\mathcal{L}_i (\mu, \theta_{i}| D_i) = \prod_j \mathcal{L}_i (\mu, \theta_i | D_{i,j})
\label{Maxlike}
\;,
\end{eqnarray}
knowing that likelihood of an individual dSph $i$, is found to be,

\begin{equation}
   \tilde{\mathcal{L}_i}(\mu,\theta_i = 
   \lbrace\alpha_i,J_i\rbrace |D_i)=\mathcal{L}_i(\mu,\theta_i|D_i)\mathcal{L}_J (J_i|J_{obs,i},\sigma_i)\;,
\end{equation}where $\mu$ encompasses the parameters of the dark matter model, i.e. the ratio of the dark matter annihilation cross section and dark matter mass squared;  $\theta_i$ accounts for the set of nuisance parameters from the LAT study ($\alpha_i$) and J-factors of the dSphs $J_i$, where $D_i$ is the gamma-ray data set. With these ingredients we perform a test statistic (TS) to obtain 95\% C.L. uppers limit on the dark matter annihilation cross section. Such bounds are derived by finding a change in the log-likelihood of $2.71/2$
with  $TS= -2 \ln ( \mathcal{L} (\mu_0,\widehat{\theta}|D)/ \mathcal{L} (\widehat{\mu},\widehat{\theta}|D))$
as described in \cite{Rolke:2004mj}. We were able to reproduce \Fermi-LAT limits using based on a joint-likelihood analysis, as demonstrated in \cite{Profumo:2016idl}. With this machinery we could investigate \Fermi-LAT sensitivity to dark matter annihilations into right-handed neutrinos as shown in Fig.\ref{limit1}-\ref{limit3}.

\section{H.E.S.S.~limits}
\label{sec:hess}
 
The H.E.S.S.~collaboration has enriched our understanding of gamma-ray emitters for energies larger than $200$~GeV, with profound impact on astrophysics and particle physics. For being an earth-based telescope, i.e.~lacking exposure to dSphs in comparison with \Fermi-LAT but gaining in effective area, the H.E.S.S.~collaboration has mostly concentrated their dark matter searches in the galactic center, which is brighter in gamma-rays than any known dSph. \\

In summary, the H.E.S.S.~collaboration derived strong limits on dark matter properties over the course of four years of observations of the galactic center which translates to 112h of data \cite{Abramowski:2011hc}. In order not to lose exposure, they divided the galactic center into source and background regions, with the former near the galactic plane where the dark matter signal is expected to be strong, and the latter
located further away from the galactic plane, where a dark matter signal is expected to be dimmer. Moreover, the source region with latitude  $|b|<0.3^\circ$ was removed to reduce the contamination by the large population of local gamma-ray sources. Anyways, afterwards the number
of events stemming from the source and background regions were obtained and can be derived using the differential flux in Figure 3 of \cite{Abramowski:2011hc} multiplied by observation time (112h), effective area provided by code gammapy \footnote{\url{https://gammapy.readthedocs.io/en/latest/} which agrees with \url{https://www.physik.hu-berlin.de/de/eephys/HESS/theses/pdfs/ArneThesis.pdf}.}, and J-factors presented in \cite{Abramowski:2011hc}. Now following the procedure described in \cite{Lefranc:2015vza}, we define likelihood functions which furnish 95\% C.L. limits on the pair {\it dark matter annihilation cross section vs mass} for a given annihilation final state and assuming a NFW dark matter density profile. We validated our procedure by comparing our results with those provided by the  H.E.S.S.~collaboration for some standard final state annihilation models such as $b\bar{b}$, finding excellent agreement as shown in \cite{Profumo:2016idl}. Recently, however, the  H.E.S.S.~collaboration has released new results using the
full statistics from 10 years (equivalent to 254 h) of galactic center observations with the
initial four telescopes of the H.E.S.S.~instrument, giving rise to a gain in sensitivity by a factor of five on the dark matter annihilation cross section for dark matter masses above $400$~GeV \cite{Abdallah:2016ygi}. The overall improvements are due to the use of a 2D binned Poisson maximum likelihood analysis which takes advantage of the spatial and
spectral characteristics of the dark matter and background expected gamma-ray emission; and the larger data set. \\

That said, knowing that for dark matter masses below $\sim 400$~GeV, \Fermi-LAT telescope offers more restrictive limits on the dark matter annihilation cross section, and the newest paper from H.E.S.S.~team does not provide much information for us to reproduce their results, we will simply rescale their limits for dark matter masses above $\sim 400$~GeV to match their current findings. This rescaling is quite reasonable as pointed out by the H.E.S.S.~collaboration in \cite{Abdallah:2016ygi}. In summary, our results based on H.E.S.S.~data rely on our determination of the H.E.S.S.~sensitivity to dark matter annihilation into right-handed neutrinos using the previous analysis by H.E.S.S.~team in \cite{Abramowski:2011hc} and the rescaling to match the updated limits.

\section{Results}
\label{sec:results}

Our results are based on the 6 years observation of dSphs by \Fermi-LAT and 10 years of data collection in the direction of the galactic center by the  H.E.S.S.~telescopes, as described in the previous sections.  In summary, we redid H.E.S.S.~and \Fermi-LAT analyses using the energy spectrum corresponding to the pair production of right-handed neutrinos, which was obtained using Pythia accounting for both 2-body and three-body decays in the context of the type-I see-saw mechanism. \\

In Figs.\ \ref{limit1}-\ref{limit3} we display upper limits on the dark matter annihilation cross section into right-handed neutrinos as a function of the dark matter mass for different right-handed neutrino masses. We present the result for each right-handed neutrino flavor separately. Hence, the electron (tau) right-handed neutrino stands for the right-handed neutrino decaying into $e\, W\, (\tau\, W)$, $\nu_e\, Z\, (\nu_\tau\, Z)$, and $\nu_\tau\, H \, (\nu_\tau\, H)$ if sufficiently heavy.

Firstly, in Fig.~\ref{limit1} we show the results for a right-handed neutrino of flavor $e$. The blue (red) curves represent the upper bounds placed on the annihilation cross section into electron right-handed neutrinos using data from the \Fermi-LAT and H.E.S.S.~telescopes. Solid, dashed and dot-dashed curves are for right-handed neutrino masses of $M_N=10$~GeV,$100$~GeV and $1$~TeV.\\

It is noticeable that \Fermi-LAT gives rise to better limits for dark matter masses below $500$~GeV, whereas H.E.S.S.~furnishes much stronger bounds for dark matter masses above $1$~TeV independently of the value used for the electron right-handed neutrino mass. For $M_{DM}=500\, {\rm GeV}$--$1\, {\rm TeV}$, the mass of the  right-handed neutrino determines which telescope performs better. This fact can be understood having in mind that the \Fermi-LAT analysis relies on photons with energies between $500$~MeV and $500$~GeV, and for dark matter particle masses at the TeV scale, some fraction of the gamma-rays are produced at energies higher than $500$~GeV, beyond \Fermi-LAT energy upper limit, weakening \Fermi-LAT sensitivity, which goes in the opposite direction of the H.E.S.S.~telescope. Therefore, these telescopes are complementary to each other for dark matter masses  between $500$~GeV-$1$~TeV.  Similar conclusions are found for the other flavors, as illustrated in figures \ref{limit2} and \ref{limit3}.  The limits for the tau  are slightly more stringent though, due to the  decay into $\tau W$, which yields a slightly stronger gamma-ray production than the other cases. In particular, we are able to exclude the canonical annihilation section of $3\times 10^{-26}{\rm \, cm^3\, s^{-1}}$ for dark matter masses below $\sim 200$~GeV.\footnote{The canonical cross section of $3\times 10^{-26}{\rm \, cm^3\, s^{-1}}$ refers to typical annihilation cross section needed to reproduce the correct relic density within the WIMP paradigm.}

\medskip

We emphasize that in the one-flavor approximation the decay branching ratio of the right-handed neutrino into a given final state is determined just by its mass. Therefore, our bounds on the dark matter annihilation cross section are not sensitive to the mixing angles, in contrast to  collider or precision studies. Consequently, gamma-ray searches introduce an orthogonal probe for right-handed neutrinos  appearing as final states from dark matter annihilation.

Our limits can be regarded as conservative, because we did not incorporate inverse Compton scattering. Inverse Compton scattering is not relevant when talking about annihilation into hadrons,  but it is relevant for annihilations $e^+e^-$ or $\mu^+\mu^-$. Notice that in our work, right-handed neutrinos may decay into $l^\pm W^\mp$, $\nu_l H$ or $\nu_l Z$. Therefore, there is always a significant hadronic contribution, rendering inverse Compton scattering not as relevant as in the case of purely leptonic annihilations such as $e^+e^-$ or $\mu^+\mu^-$. Notice from our figure 4, that changing the final state from $e$ to $\mu $ yields basically no change to our results, showing that the hadronic contribution is dominant. Moreover, inverse Compton scattering is particularly not important in the case for dSphs  due to low interstellar radiation field and unknown diffusion. However, in the H.E.S.S. analysis, the galactic center is the target. Thus, inverse Compton	 can indeed be important, but is subject to large uncertainties due to our lack of knowledge regarding the diffusion coefficient. Therefore, the addition of more unknown  quantities to our work because of two possible final states relevant only to Fermi-LAT study is not well motivated. Nevertheless, keep in mind that depending on the diffusion coefficient used, stronger limits on annihilation cross section could be achieved.

\section{Conclusions}
\label{sec:con}

In the light of the importance of right-handed neutrinos in several model building endeavors, and their appearance as possible final states from dark matter annihilation, we discussed a probe for dark matter annihilations into right-handed neutrinos using gamma-ray  data. Analyzing \Fermi-LAT 6-year data  of dwarf spheroidal galaxies, and  H.E.S.S.~10-year  data  of the galactic center region, we placed stringent bounds on dark matter annihilations into right-handed neutrinos. In particular, we ruled out the thermal annihilation cross section  for dark matter masses below 200 GeV. This limit concretely shows that the search for gamma-rays from dark matter annihilation brings out a test to dark matter models where right-handed neutrino arise as possible annihilation final states. 

\section*{Acknowledgments}

We thank the Fermi-LAT Collaboration for the public data and tools used in this work. We are also grateful to the PPPC4DMID team for making their code publicly available. The authors warmly thank Werner Rodejohann for his collaboration in early stages of the project. FSQ thanks Torsten Bringmann for comments and University of Olso for the hospitality  during which this project was partly carried on. The authors are grateful to William Shepherd, Evgeny Akhmedov and Alexey Smirnov for discussions. C.W.~was supported by NWO through one Vidi grant. M.C.~was supported by the IMPRS-PTFS.

\bibliographystyle{JHEPfixed}
\bibliography{darkmatter}

\end{document}